\documentstyle[prb,aps]{revtex}
\begin{document}
\title{Band structures and optical properties of
Ga$_{1-x}$In$_x$As quantum wires grown by strain-induced lateral ordering}

\author{Liang-Xin Li and Yia-Chung Chang}
\address{Department of Physics and Materials Research Laboratory 
\\ University of Illinois at Urbana-Champaign, Urbana, Illinois 61801}

\date{\today}
\maketitle

\begin{abstract}
Band structures and optical matrix elements of strained 
multiple quantum-wires (QWR's) are investigated theoretically via the
effective bond-orbital model, which takes into account the effects of 
valence-band anisotropy and the band mixing. 
In particular, the Ga$_{1-x}$In$_x$As QWR's grown by 
strain-induced lateral ordering (SILO) are considered.
Recently, long wavelength Ga$_{1-x}$In$_x$As QWR lasers have been 
fabricated via a single step molecular beam epitaxy technique which uses the 
SILO process.[1] Low threshold current and high optical anisotropy 
have been achieved. Multi-axial strains [combinations of biaxial
strains in the (001) and (110) planes] for QWR's are considered, 
Our calculated anisotropy in optical matrix elements 
(for light polarized parallel versus perpendicular to the QWR's axis) is in 
good agreement with experiment. We also find that the strain 
tends to increase the quantum confinement and enhance the anisotropy of 
the optical transitions.
 \end{abstract}

\mbox{}

\mbox{}

\newpage

\section{Introduction}

$Ga_xIn_{1-x}As$ is one of the most important ternary III-V 
compound semiconductors. Its bandgap covers both the 1.3 and 1.55 $\mu m$ 
range, which are the preferred wavelengths in long distance fiber 
communications[1,2]. However, long-wavelength photonic devices based on 
lattice-matched $Ga_{0.47}In_{0,53}$As/InP heterostructures suffer 
from strong Auger recombination and intervalence band absorption 
processes[7-10].
Recently, to improve the performance of long-wavelength semiconductor 
lasers, long wavelength($\sim 1.55 {\mu}$m) $Ga_xIn_{1-x}As$ quantum-wire 
(QWR) lasers have been grown by a single step molecular beam 
epitaxy technique[1]. It is found that the QWR laser structures are a 
promising choice because of the many predicted benefits, such as higher gain, 
reduced temperature sensitivity, higher modulation bandwidths, and narrower 
spectral linewidths[2]. Short-period superlattices (SPS)  in the direction
perpendicular to the growth direction can be formed  
via the strain-induced lateral-layer ordering 
(SILO) process[1,2]. A quantum wire heterostructure can then be 
created by simply utilizing this SPS structure as the quantum well region in a 
conventional quantum heterostructure. Besides the self-assembled lateral 
ordering, it is believed that the strain also plays a key role[2] 
in the temperature stability and high anisotropy for the QWR laser structure. 
Recent studies indicate that the use 
of the strained-layer quantum wire heterostructure has advantages of 
high-quality interfaces and band-gap tuning which are 
independent of the lattice constant of the 
constituent materials[3,4,6]. Furthermore, much work has been undertaken which 
predicts that by using strained-layer superlattice to form the active 
region of a 
quantum-wire laser, the threshold current can be decreased by one order of 
magnitude, and the optical loss due to intervalence-band absorption and  Auger 
recombination will also be greatly reduced[1,2,7-10]. Also, the temperature 
sensitivity 
is reduced by a order of magnitude compared with strain-free structures[2]. 
A typical 
temperature sensitivity of the lasing wavelength is($\sim 5 \AA/{}^0C$) for 
the usual 
GaInAs/InP lasers. By using a distributed-feedback structure, the temperature 
dependence of the lasing wavelength is reduced to $1\AA/{}^0C$[2]. 
With this strained 
GaInAs QWR, the dependence is smaller than $0.1\AA/{}^0C$[2]. It should be an 
important improvement on the current technology in fabricating the 
long wavelength lasers for fiber communication. 
In this paper we study the effects of multi-axial strain on
the electronic and optical properties of the QWR structures grown via 
the SILO process. Through this study, we can gain a better 
understanding of the strain engineering of  
QWR structures suitable for the application of fiber-optical communication. 

   In the experiment performed by  Chou et al[1], the QWR active region is created 
$\it{in~ situ~}$ by the SILO process within the $(GaAs)_2/(InAs)_{2.2}$ SPS regions. 
The SILO process generates a strong Ga/In lateral 
composition modulation and creates In-rich $Ga_xIn_{1-x}As$ lateral QWs in the [110] 
direction. By sandwiching the In composition modulated layer (with $x$ varying 
from 0.3 to 0.7) 
between $Al_{0.24}Ga_{0.24}In_{0.52}As$ barrier layers, a strained QWR 
heterostructure is formed[1,2]. The model QWR structure considered in the 
present paper is depicted in Figure.1. 
The central region consists of the In composition modulated $Ga_xIn_{1-x}As$ 
layer (A/B/A/B...), 
in which the A strip is Ga-rich while the B strip is In-rich. The lateral 
period is $L_2=L_2^A+L_2^B$.
In our calculation, we consider $L_2^A=L_2^B = 50\AA$ and 
$L_{\perp}^B=L_{\perp}^W=50\AA$.
We consider two cases of composition modulation. In case one, the 
Ga composition changes abruptly
with $x=0.6 (or 0.7)$ in strip A and $x=0.4 (or 0.3)$ in strip B. 
In case two, $x$ varies 
sinusoidally from 0.6 (or 0.7) at the 
center of strip A to 0.4 (or 0.3) at the center of strip B 
(with $x=0.5$ at the border between A and B strips).

\section{Theoretical Approach}

\ The method used in this paper for calculating the strained quantum well band 
structure is based 
on the effective bond-orbital model (EBOM). A detailed description of this 
method without 
the effect of strain has been published elsewhere[5]. Here we give a brief 
account of the method 
and discuss the effect of strains. The
effective bond-orbital model employed here is a tight-binding-like model which includes 
nearest-neighbor interactions among bond orbital residing on an fcc lattice. 
The version of the 
EBOM used in this calculations describes the coupling between the upper four 
spin-3/2 valence 
bands and the lowest two s-like conduction bands. For our QWR system, the 
split-off bands are 
ignored since the spin-orbital splitting is large. A bond orbital is defined 
to be the proper linear combination of the two atomic orbitals within a 
zinc-blende crystal which best describes the 
states near the center of the Brillouin zone. The parameters that appear in 
the theory are given 
by a correspondence which is made by requiring that the Hamiltonian in the 
bond-orbital basis, 
when written for the bulk material and expanded to the second order in 
${\bf k}$, agree with the 
Luttinger-Kohn expression. For interactions across heterojunctions, we take 
the average of the 
matrix elements for the two bulk materials. EBOM has been successfully 
applied to the 
calculation of electronic states of superlattices[5], strained-layer 
superlattices [6,19], semiconductor
quantum wires[15] and dots[17], and impurity levels in quantum dots[17] 
and quantum wells[17]. 
Here, we apply it to this multi-axially strained QWR structures.

  We solve the QWR electronic states according to a procedure similar to 
that described in [15]. 
We first calculate the band structure of a superlattice(SL) with M bilayers of
$Ga_{x}In_{1-x}As$ (the well material) and N bilayers of  
$Al_{0.24}Ga_{0.24}In_{0.52}As$
(the barrier material) using the slab method. The superlattice period is
$L_{\perp}=L^B_{\perp}+ L^W_{\perp} $= (M+N)a/2, where a is the lattice 
constant. For a 
sufficiently large value of the barrier thickness (we used $ L^B_{\perp} = 
50\AA$) the QW's 
are essentially decoupled and we ignore the dependence on the wave number 
in the growth 
direction, taking $k_z = 0$. The SL eigenstates with $k_z =0$ are denoted 
by $|k_1, \bar k_2, n>$, i.e,
\begin{equation}
   H^{SL}|k_1,\bar k_2,n>= E_n(k_1,\bar k_2)|k_1,\bar k_2,n>
\end{equation}
where $H^{SL}$ is the Hamiltonian for the  
$Ga_{y}In_{1-y}As$/$Al_{0.24}Ga_{0.24}In_{0.52}As$ SL ($y=0.6 or 0.7$)
,where $k_1$ and $k_2$ are the two components of the SL in-plane wave vector 
along and 
perpendicular to the QWR axis, respectively. $k_1$ is a good quantum number 
for the QWR array, 
while $\bar k_2$ is not. The QWR wave function is a linear combination of SL wave 
functions with 
the same $k_1$ but with values of $\bar k_2$ differing by 2$\pi/L_2$. 
We define $k_2$  to be  
the wave number of the QWR array 
in the [110] direction, and we can write $\bar k_2= k_2 + 2\pi\zeta/L_2$ 
with $\zeta=0,1,2,...,$m, 
where $m = L_2/a_2$ with $a_2= a/\sqrt{2}$. 
The QWR Hamiltonian is $H^{QWR}=H^{SL} + H'$, where 
$H'= H_{Ga_{x}In_{1-x}As}-H_{Ga_{y}In_{1-y}As}$  
is treated as a perturbation. The Hamiltonian matrix for QWR
\begin{equation}
M(n,\zeta|n', \zeta') = 
<k_1,k_2+ 2\pi\zeta'/L_2|H^{QWR}|k_1,k_2+ 2\pi\zeta/L_2,n>
\end{equation}

\noindent is then diagonalized in the basis consisting of SL eigenstates. 
The basis is truncated
to reduce the computer time, thus making the method a variational one. 
The subbands 
closest to the band edge are converged to within 0.1 mev. The size of the 
Hamiltonian matrix, originally given by the number of sites in a QWR 
supercell 
times the number of coupled bands, is now reduced to the number of SL states 
necessary to give accurate results for the 
QWR of concern. In our case, this number needed to get well converged results 
for the
two uppermost pair of valence subbands is around 200. However, we have used 
$\sim 800$ basis states 
in order to ensure the accuracy of the deeper lying sates for all 
calculations 
in this present paper. 
The effect of lateral intermixing of Ga and In is accounted for by using 
the virtual-crystal approximation.
Bond-orbital parameters for the inter-diffused materials are 
obtained by linearly interpolating 
between the values of the parameters for the pure materials. 
The optical parameters in these 
expressions are determined by requiring that the optical 
matrix elements between bulk states 
obtained by EBOM be identical to those obtained in the ${\bf k\cdot p}$ 
theory up to second 
order in ${\bf k}$[15].

    The effect of strain is included by adding a strain Hamiltonian 
$H^{st}$ to the EBOM effective
Hamiltonian[6,10,16].  The matrix elements 
of $H^{st}$ in the bond-orbital basis can be 
obtained by the deformation-potential theory of Bir and Pikus[16]. 
Here, we consider the combination
of two bi-axial strains, one in the (001) plane 
(due to the lattice mismatch at the interfaces
between GaInAs and AlGaInAs) and the other in the (110) plane 
(due to the Ga composition modulation ). 
The resulting strains in the SL layers consist 
of both hydrostatic and uniaxial components. For case 1 (Ga composition
changes abruptly from region A to region B), the lattice constant 
of the strained layers in the [001] direction  
($\tilde a_{001}$) is given by minimizing the free energy of 
the fcc system due to strain[11]:
\begin{eqnarray}
F= \frac{1}{2}C_{11}^A(\epsilon_{11A}^2 + \epsilon_{22A}^2+\epsilon_{33A}^2)
L_2^A~~~~~~~~~~~~~~~~~~~\\\nonumber
~~~~~+ C^A_{12}(\epsilon_{11A}\epsilon_{22A}+\epsilon_{22A}\epsilon_{33A}
+\epsilon_{11A}\epsilon_{33A})L_2^A~~~\\\nonumber
+\frac{1}{2}C^B_{11}(\epsilon_{11B}^2 + \epsilon_{22B}^2+\epsilon_{33B}^2)
L_2^B~~~~~~~~~~~~~~~~~\\\nonumber
+ C^B_{12}(\epsilon_{11B}\epsilon_{22B}+\epsilon_{22B}\epsilon_{33B}
+\epsilon_{11B}\epsilon_{33B})L_2^B~~,
\end{eqnarray}
where $L_2^A$ ($L_2^B$) is the lateral layer thickness and $\epsilon_{ijA} 
(\epsilon_{ijB})$ is the strain tensor in the
Ga-rich (In-rich) region. $C^A$'s ($C^B$'s) are the elastic 
constants for the Ga-rich (In-rich) materials.
Here we have used a rotated Cartesian coordinates in
which $x'=[1\bar 1 0], y'=[110],$ and $z'=[001]$.
A constraint $\tilde{a}^A_{\|}=\tilde{a}^B_{\|}=a_{InP}=a_{AlInGaAs}$ 
has been imposed to keep the  $Al_{0.24}Ga_{0.24}In_{0.52}As$ barrier 
strain free, since it is lattice-matched to the InP substrate[1,2],
where $\tilde{a}^\alpha_{\|}$ is the in-plane strained lattice constant 
(perpendicular to [001] axis)
for material $\alpha$. The multi-axial strain caused by the lattice-mismatch
in both (001) and (110) planes is simply
\begin{equation}
\epsilon_{11\alpha} = \epsilon_{22\alpha} =
\frac{\tilde{a}^\alpha_{\|}-a_\alpha}{a_\alpha},
\end{equation}
where $a_A$ ($a_B$) is the unstrained lattice constant of $Ga_{x}In_{1-x}As$
($Ga_{1-x}In_{x}As$) ($x=0.6$ for case 1) 
and 
\begin{equation}
\epsilon_{33\alpha} = \frac{\tilde{a}_{001}-a_\alpha}{a_\alpha}
\end{equation}
with $\epsilon^\alpha_{ij} = 0$ for $i \ne j$.

The minimization procedure leads to 
\begin{equation}
\tilde{a}_{001}=-\frac{1}{2}\frac{ C_{11}^A L_2^A/a_A^2 + 
C_{11}^B L_2^B/a_B^2}{(C_{12}^A \epsilon_{\|}^A L_2^A/a_A+C_{12}^B 
\epsilon_{\|}^B L_2^B/a_B)+(C_{11}^A L_2^A/a_A+C_{11}^B L_2^B/a_B)},
\end{equation}
which is the strained lattice constant in the [001] direction.

Due to symmetry, the multi-axial strain tensor is diagonal 
in the $x'y'z'$ coordinates, which leads to a diagonal
strain Hamiltonian 
\begin{equation}
H^{st} = \left( \begin{array}{ccc} 
                                        -\Delta V_H + D_1 & 0 & 0 \\
                                          0 & -\Delta V_H + D_2  & 0\\
                                          0 & 0 & -\Delta V_H + D_3
               \end{array} \right),
\end{equation}
where 
$$
\Delta V_H = (a_1+a_2)(\epsilon_{11}+\epsilon_{22}+\epsilon_{33}),~~\\
 D_1 = b(2\epsilon_{11}-\epsilon_{22}-\epsilon_{33}),~~\\
 D_2 = b(2\epsilon_{22}-\epsilon_{11}-\epsilon_{33}),~~\\
 D_3 = b(2\epsilon_{33}-\epsilon_{22}-\epsilon_{11}),\\
$$
The strain potential on the s states is given by[6]
$$\Delta V_c = c_1(\epsilon_{11}+\epsilon_{22}+\epsilon_{33}),$$
The strain Hamiltonian in the bond-orbital basis $|JM>$ can be easily 
found by using the coupling constants[5], i.e, 
\begin{equation}
<JM|H^{st}|J'M'>=\sum_{\alpha, \alpha',\sigma}C(\alpha, \sigma;J,M)^*
C(\alpha', \sigma;J',M')H^{st}_{\alpha\alpha'}
\end{equation}
The elastic constants $C_{12}$ and $C_{11}$ for GaAs, InAs and AlAs can be found
in Ref. [12,18]. The deformation potentials$a_1,~a_2,~b,~c_1$ can be
found in Ref.[20,21]. The linear interpolation
and virtual crystal approximation is used to obtain the corresponding 
parameters for the GaInAs and AlGaInAs. 

After rotating back to the original Cartesian coordinates, the final 
results for the strain Hamiltonian in the bond-orbital basis $|JM>$ is
\begin{equation}
 H^{st} =\left( \begin{array}{cccccc} 
               \Delta V_c & 0 & 0 & 0 & 0 \\
                0 & \Delta V_c & 0 & 0 &0 &0 \\
                0 & 0 & -V_H + A& 0 &B/\sqrt{12} &0  \\
                0 & 0 & 0 & -V_H + C & 0 & B/\sqrt{12} \\
                0 & 0 & B/\sqrt{12} & 0 & -V_H + C & 0 \\
                0 & 0 & 0 & B/\sqrt{12} & 0 & -V_H + A \\
                          \end{array} \right)
\end{equation}
with
$$
A=\frac{D_3+0.5 (D_1+D_2)}{2},~~
B=\frac{(D_1+D_2)}{2}-D_3,~~
C=\frac{2.5 (D_1+D_2)+D_3}{6}.
$$
One can see non-diagonal terms arise due to the 
rotation, which will cause
additional mixing and optic transitions between different valence bands.  

	The above strain Hamiltonian is derived for case one only, where
the Ga composition ($x$) changes abruptly from region A to region B. 
For case two, $x$ varies continuously from $x_m$ to $1-x_m$ ($x_m=0.6 or 0.7$).
In this case, we shall 
first calculate the strain Hamiltonian for the abrupt case 
to get $H^{st}_{x_m}$ and $H^{st}_{1-x_m}$, using Eq. (8).
We then obtain the strain Hamiltonian at any atomic layer with Ga composition
$x$ via a linear interpolation between $H^{st}_{x_m}$ and $H^{st}_{1-x_m}$.
This is consistent in spirit with the virtual crystal approximation we have 
used for obtaining the interaction parameters in EBOM. 

\section{Results and discussions}

In Figure 2, we show
a schematic diagram indicating the alignment between band edges 
of the constituent materials for QWR in case 1 (square profile)
with and without the effect of strain. 
All energies are measured with respect to the bulk InAs valence band edge. 
This diagram is useful in understanding the 
quantum confinement effect on the QWRs considered here.
From Figure 2, we notice that the strain [in particular, the one in the (110)
plane] has a significant effect on the band alignment. Take Figure 2(a) for example,
without strain (solid
lines) the conduction (valence) band of Ga$_{0.4}$In$_{0.6}$As is below (above)
that of Ga$_{0.6}$In$_{0.4}$As by 247 meV (14 meV).
With strain (dotted lines), the situation is reversed and
the conduction (valence) band of Ga$_{0.4}$In$_{0.6}$As is above (below)
that of Ga$_{0.6}$In$_{0.4}$As by 312 meV (76 meV); thus, both electrons and
holes will be confined in the Ga-rich region of the QWR. We found that
about 80 \% of these changes are caused by the strain in the (110) plane, since
there is a stronger lattice mismatch between Ga$_{0.6}$In$_{0.4}$As 
and Ga$_{0.4}$In$_{0.6}$As compared with that between Ga$_{0.6}$In$_{0.4}$As and 
Al$_{0.24}$Ga$_{0.24}$In$_{0.52}$As (or InP).

The strain also causes a splitting between the heavy-hole (HH)  and light-hole (LH)
bands with the LH band lying above the HH band. However, the splitting is rather
small, about 3 meV in Ga$_{0.6}$In$_{0.4}$As. When the confinement effect
due to the barrier material (Al$_{0.24}$Ga$_{0.24}$In$_{0.52}$As) is included,
the HH band again lies above the LH band, due the difference in the effective
masses along the growth direction ($z$). The dash-dotted lines in Figure 2(a) 
indicate the superlattice band edges of 50\AA Ga$_{0.6}$In$_{0.4}$As 
(or Ga$_{0.4}$In$_{0.6}$As)
sandwiched between 50\AA Al$_{0.24}$Ga$_{0.24}$In$_{0.52}$As barriers. 
The difference in superlattice band edges (dash-dotted lines) between 
Ga$_{0.6}$In$_{0.4}$As and Ga$_{0.4}$In$_{0.6}$As determines the degree of
lateral quantum confinement in the QWR. For the present case, the conduction-band
offset is 252 meV and the valence-band (for HH only) offset is 52 meV, as
far as the lateral confinement is concerned. Both offsets are large enough
to give rise to strong lateral confinement for electrons and holes in
the Ga-rich region.

  Figure 3 shows the near zone-center valence subband structures of 
square-shaped QWRs with the multi-axial strain with (a) Ga composition ($x$) 
changing abruptly from 0.6 in region A to 0.4 in region B and (b)Ga composition 
($x$) changing abruptly from 0.7 in region A to 0.3 in region B.                     
To compare with experiment[1,2], we 
choose material parameters appropriate for temperature at 77K[12]. 
All subbands are two-fold degenerate due to the Kramer's degeneracy and
they are labeled according to the characters of their underlying Bloch
functions at the zone-center: HH for heavy hole and LH for light-hole 
(only the confined subbands are labeled).      
The conduction subbands are approximately 
parabolic as usual with a zone-center subband minimum equal to
826 meV in (a) and 707 meV in (b) (not shown in the figure). 
This gives an energy gap 860 meV for the (x=0.6/0.4)
QWR array and 708 meV for the (x=0.7/0.3) QWR array . 

The observed C1-HH1 excitonic transition is at 735 meV (or 1.65$\mu m$)
for the QWR[1,2] at 77K. 
In comparing the band gaps with the experiment, one should also take into 
account the exciton binding energy which is around
20 meV for this size of QWR.
Thus, the theoretical result for the (x=0.7/0.3) QWR array is in closer 
agreement with experiment, but about 50 meV too low. 

   Comparing the band structures in both $k_1$ ($[1\bar 10]$) and $k_2$ ($[110]$) 
directions, we noticed an apparent anisotropy in the energy dispersion.
The dispersions in the $k_2$ direction for the five uppermost (confined) valence
bands are rather small, indicating strong lateral confinement.
We observe strong anti-crossing effect between the HH1/HH2 and LH1 subband at
$k_1$ near $0.02 \AA^{-1}$ similar to what happens in a quantum well
\cite{Sanders}. 
Here the HH2 subband of the QWR corresponds to the [110] zone-folded  part of the 
HH1 subband of the quantum well ( the envelope function
has odd parity in the [110] direction but even parity in the $[1\bar 10]$ direction).

	The subbands in Figure 3(b) have less dispersion in the $k_2$ direction
compared with those in Figure 3(a) as a result of stronger lateral confinement.
This is caused by the larger band discontinuities in the x=0.3/0.7 case (versus
the x=0.4/0.6 case) as can be seen by comparing band alignments shown in Figure 2.

     Figure 4 shows the near zone-center valence subband structures of 
a QWR with a sinusoidal lateral modulation with Ga composition ($x$)
varying as the position ($y'$) in the [110] direction between two extreme values
$x_m$ and $1-x_m$, i.e. $$x(y')=0.5 + (0.5-x_m) sin(2\pi y'/L_2)$$ for
$x_m=0.4$ and (b) $x_m=0.3$.  
In this case, the Hamiltonian including the multi-axial strain is 
calculated with a linear extroplation between the Hamiltonians
for the maximum and minimum Ga composition, i.e.
$$
H_{Ga_{x}In_{1-x}As}^{st}=f H_{Ga_{x_m}In_{1-x_m}As} + 
(1-f) H_{Ga_{1-x_m}In_{x_m}As},
$$
where $x_m=0.4$ or 0.3 and $f$ is determined by comparing the Ga composition 
on both sides of the equation
$$ 
x = x_m f + (1-x_m) (1-f),
$$
or $f= (1-x_m-x)/(1-2x_m)$.

The conduction band minimum (not shown) is 850 meV (757 meV) for the
case $x_m=0.4$ ($x_m=0.3$), which corresponds to
a band gap of 889 meV (767 meV). Subtracting the exciton binding energy 
($\sim$ 20 meV)
from the band gap, we found that the $x_m=0.3$ case is in fairly good agreement 
with the observed excitonic transition energy of 735 meV.
The band gaps of QWR's with the sinusoidal profile are consistently larger
(by about 30-50 meV)
than the corresponding QWR's with the square-profile (with the same extreme 
values of Ga composition). 
The sinusoidal variation in Ga composition gives rise to more energy, 
since the well region contains more materials with band gap higher 
than the minimum in comparison with the square-profile.
Furthermore, the lateral strain is reduced (less mismatch on the average) 
which also tend to increase the band gap.

	Comparing Figure 4 with Figure 3, we notice that the spacing
between HH1 and HH2 is substantially larger for the sinusoidal profile
than for the square profile. This can be understood by the following argument.
The envelope function for the HH2 state is more spread out than that for
the HH1 state, thus its energy is increased more in the sinusoidal profile
(with higher probability being in materials with higher band gap)
compared with the square profile.  

	Next we discuss the optical properties of QWR's. Since the QWR states
can be qualitatively viewed as the zone-folded states of the superlattice (SL) 
states (the zeroth-order states) via the lateral confinement, it is instructive 
to examine the optical matrix elements of the SL case. Figure 5 shows the
squared optical matrix elements of the (50\AA/50\AA)
Ga$_{0.7}$In$_{0.3}$As/Al$_{0.24}$Ga$_{0.24}$In$_{0.52}$As superlattice
for the HH1-C1 and LH1-C1 transitions versus the wave vector $k_2$. 
The solid and dashed lines are for
the polarization vector along the $k_1$ ($[1\bar 1 0]$) and $k_2$ ([110])
directions, respectively. We note that the optical matrix elements are isotropic
in the $x$-$y$ plane at the zone-center, while they become anisotropic at
finite values of $k_2$. At the zone center, the HH state consists of  
Bloch states with atomic character $(x'+iy')\uparrow$ ($|J,M>=|3/2, 3/2>$), 
while the LH state  consists of Bloch states with atomic character 
$(x'-iy')\uparrow+2z\downarrow$ ($|J,M>=|3/2,-1/2>$); 
thus, the corresponding optical transitions 
(to an s-like conduction band state) are isotropic in the $x$-$y$ plane.
Here $x' (y')$ is the coordinate along the $k_1 (k_2)$ direction.
For finite $k_2$, the HH and LH characters are mixed, with the HH state 
consisting of more $x'$-character than $y'$-character, thus in favor of
the polarization vector parallel to the $k_1$ direction. Note that the HH (LH) 
band tends to have an atomic character with polarization perpendicular 
(parallel) to the direction of wave vector. This is a direct consequence
of the fact that the ($pp\pi$) interaction is weaker that the ($pp\sigma$) 
interaction in a tight-binding model.
Since QWR states are derived from the SL states with
finite values of $k_2$, we expect the optical matrix elements of QWR
to be anisotropic in the $x$-$y$ plane as well.

Figures 6 and 7 show the squared optical matrix elements versus wave vectors 
of the corresponding QWR's considered in Figures 3 and 4, respectively.
respectively. The solid and dashed lines are for
the polarization vector along the $k_1$ ($[1\bar 1 0]$) (parallel to
the wire) and $k_2$ ([110]) (perpendicular to the wire) directions.  
The optical matrix elements together with the subband structures  discussed 
above provide the essential ingredients for understanding the optical 
transitions observed in the photoluminescence (PL) measurements. 
In all figures the optical transitions considered are
from the topmost three valence subbands to the lowest conduction subband: 
HH1-C1, HH2-C1, and LH1-C1. These curves were obtained by summing 
the contributions of two degenerate subbands (due to Kramer's degeneracy)
for the initial and final states.
To compare with experimental (PL) results, we concentrate on the HH1-C1 
transition. 

From Figures 6 and 7,  we found for the parallel polarization (solid lines), 
the squared optical matrix element for the HH1-C1 transition has a maximum at 
the zone center with a value near 25 eV and remains close to this maximum 
value for all finite $k_2$ and with $k_1=0$. For the
perpendicular polarization, the HH1-C1 transition is very small for
finite $k_2$ and $k_1=0$.  This means that the wave function of the HH1 state 
has mostly $x'$ character (and some $z$ character). 
Thus, we conclude that the strong 
lateral confinement forces the wave function of the HH1 state with $k_1=0$ to 
change from the $x'+iy'$ character (in the SL case) into mostly $x'$ character.
At finite $k_1$, the character of the HH1 state gradually changes back
toward  the $x'+iy'$ character and the optical transition becomes almost 
isotropic when $k_1$ is comparable to zone-boundary value of 
$k_2 (\pi/L_2)$, and finally turns into mostly $y'$ character as
$k_1$ becomes much larger than $(\pi/L_2)$. The
anti-crossing behavior of the HH1/HH2 subbands with the LH1 subband 
further complicates the whole picture at $k_1 \approx 2 \AA^{-1}$.

The HH2-C1 transition has zero optical strength at the zone center for
both polarizations. This is expected, since the HH2 state has odd parity in the
envelope function, which leads to forbidden transition at the zone center.
The symmetry restriction is relaxed as the wave vector deviate from zero.
The LH1-C1 transition has large optical strength (around 17 eV)
for the perpendicular
polarization and very weak strength (around 1 eV) for the parallel 
polarization at the zone center.
This indicates that the LH1 states consists of mostly $y'$ character.
So the strong lateral confinement forces the the wave function of the LH1 state 
with $k_1=0$ to change from the $x'-iy'$ character (in the SL case) 
into mostly $y'$ character. In other words, the HH1 state (with pure 
$x'+iy'$ character) and LH1 state (with pure $x'-iy'$ character) in the SL case
are mixed thoroughly by the lateral confinement in QWR to produce
a predominantly $x'$ state (HH1) and a predominantly $y'$ (LH1) state.
Note that all thse states consist of appreciable $z$ character, which
will appear in the optical transition with $z$ polarization.

To calculate the anisotropy, we integrate the squared optical matrix 
element over range of $k_1$ corresponding to the spread of exciton 
envelope function in the $k_1$ space. The exciton envelope funtion
is obtained by solving the 1D Schr\"{o}ding equation for the exciton in
the effective-mass approximation
\begin{equation} 
[ - \frac {\hbar^2} {2\mu} (\frac  {\partial} {\partial x'})^2
 + V_X(x')]F(x')=E_X F(x'), \end{equation}
where $\mu$ is the exciton reduced mass ($\approx 0.037 m_0$)
and $V_X(x')$ is the effective 1D exciton potential given by\cite{Sanders1}
\[ V_X(x')=\frac {e^2}{\epsilon_0 |x'|}(1-e^{-\beta|x'|}). \]
The parameter $\beta$ is obtained by extropolation from the values given in
Table V of Ref. \cite{Sanders1}. For a $50\AA$ quantum wire, we obtain 
$\beta=0.07 \AA^{-1}$. Eq. (10) is solved numerically, and the exciton
envelope function in $k_1$ space is obtained via the Fourier transform of
$F(x')$. The exciton binding energy obtained is 23 meV.
We found that the ratio of the averaged  
optical strength for the HH1-C1 transition for the parallel to perpendicular 
component of the polarization vector is 2.65 (4.04) for the $x=0.4/0.6$ 
($x=0.3/0.7$)
QWR with square profile, and 2.86 (7.19) for the $x_m=0.4$ ($x_m=0.3$)
QWR with sinusoidal profile.  The $x_m=0.3$ case shows stronger
optical anisotropy than the $x_m=0.4$ case, indicating that the stronger 
lateral confinement leads to stronger optical anisotropy as expected.
Experimentally, the optical anisotropy is found to be around 2-4. 
So, our results are consistent with experiment. 

\section{Conclusion}

We have calculated the band structures and optical matrix elements 
for the strained QWR grown by the SILO method. The effect of multi-axial strain 
on the valence subbands and optical matrix elements is discussed.  
Our theoretical studies provide the explanation of the anisotropy in optical 
matrix elements of these QWR's observed experimentally.
We find that the biaxial strain due to the lattice mismatch between the
Ga-rich and In-rich regions is most dominant. It 
tends to increase the lateral confinement in the QWR and enhances the 
anisotropy of the optical transitions which may be useful 
for certain applications in optical communication. We
also calculated the effect of lateral composition modulation on the band 
structures and optical properties and find that it increases the band gaps
and reduces the optical anisotropy. From the above discussions, we can use the
theoretical predictions to guide the  engineering design of QWR optical devices.
The temperature effect will be incorporated in our future studies
in order to understand the temperature stability of the optical transitions of
QWR's grown by the SILO method.

\vspace{1cm}
\leftline{\bf ACKNOWLEDGEMENTS}
\vspace{1ex}
  This work was supported in part by the National Science Foundation (NSF) 
under Grant No. NSF-ECS96-17153. We would like to thank K. Y. Cheng
and D. E. Wohlert for fruitful discussions
and for providing us with the detailed experimental data
of the QWR structures considered here.

\newpage
\vspace{2cm}
Figure Captions

\vspace{1ex}

\noindent Fig. 1. Schematic sketch of the QWR array fabricated in Refs. [1,2]. 
The QWR axis 
lies in the $[1\bar 10]$ direction. A and B label the Ga-rich and In-rich 
strips, respectively. 

\noindent Fig. 2. Schematic diagram indicating the alignment between band edges 
of the constiuent materials for QWR in case 1 (square profile)
with and without the effect of strain. Solid lines are for unstrained bulk,
dotted lines are for bulk under multi-axial strain appropriate for the
present QWR, and dash-dotted lines are for (50\AA/50\AA) 
GaInAs/AlGaInAs superlattice  including the multi-axial strain. 

\noindent Fig. 3. Valance subband structures for QWR's with square profile
for (a) Ga-composition $x=0.6$ in Ga-rich and $x=0.4$ in In-rich region and
(b) $x=0.7$ in Ga-rich and $x=0.3$ in In-rich region.

\noindent Fig. 4. Valance subband structures for QWR's with 
sinusodial profile with $x$ ranging from $x_m$ to $1-x_m$ 
for (a) $x_m=0.6$ and (b) $x_m=0.7$.

\noindent Fig. 5. Squared optical matrix elements for transitions from HH1 
and LH1 to the first conduction subband of the (50\AA/50\AA)
Ga$_{0.7}$In$_{0.3}$As/Al$_{0.24}$Ga$_{0.24}$In$_{0.52}$As superlattice
for light polarized parallel (solid), perpendicular (dashed) 
to the QWR axis and z component (dotted).            

\noindent Fig. 6. Squared optical matrix elements for transitions from 
the top three valence subbands to the first conduction subband
for light polarized parallel (solid) and perpendicular (dashed) 
to the QWR axis for QWR's considered in Fig. 3.
                  
\noindent Fig. 7. Squared optical matrix elements for transitions from 
the top three valence subbands to the first conduction subband
for light polarized parallel (solid) and perpendicular (dashed) 
to the QWR axis for QWR's considered in Fig. 4.

\newpage

\centerline{\bf Fig.1}
\vspace{1cm}

\begin{picture}(100,220)(-80,20)
\put(40, 220){\line(1,0){300}}
\put(160,190){\makebox(0,0){$Al_{o.24}Ga_{0.24}In_{0.52}As$}}
\put(40, 160){\line(1,0){300}}
\put(40,140){\makebox(0,0){$\cdot\cdot\cdot$}}
\put(60, 160){\line(0,-1){60}}
\put(70,140){\makebox(0,0){~~~B}}
\put(100, 160){\line(0,-1){60}}
\put(110,140){\makebox(0,0){~~~A}}
\put(265, 140){\vector(0,1){20}}
\put(265, 130){\makebox(0,0){$L_{\perp}^W$}}
\put(265, 120){\vector(0,-1){20}}
\put(265, 75){\vector(0,1){25}}
\put(265, 70){\makebox(0,0){$L_{\perp}^B$}}
\put(265, 65){\vector(0,-1){25}}
\put(140, 160){\line(0,-1){60}}
\put(150,140){\makebox(0,0){~~~B}}
\put(180, 160){\line(0,-1){60}}
\put(190,140){\makebox(0,0){~~~A}}
\put(220, 160){\line(0,-1){60}}
\put(230,140){\makebox(0,0){~~~$\cdot\cdot\cdot$}}
\put(260, 160){\line(0,-1){60}}
\put(300, 160){\line(0,-1){60}}
\put(40, 60){\makebox(0,0){$\bigodot$}}
\put(45, 60){\vector(1,0){20}}
\put(40, 65){\vector(0,1){20}}
\put(42,48){\makebox(0,0){[1$\bar{1}0]~\vec{k_1}$}}
\put(76,60){\makebox(0,0){[110]}}
\put(96,60){\makebox(0,0){$\vec{k_2}$}}
\put(48,88){\makebox(0,0){[001]}}
\put(40, 100){\line(1,0){300}}
\put(220, 100){\line(1,0){40}}
\put(40, 40 ){\line(1,0){300}}
\put(180,60){\makebox(0,0){$Al_{o.24}Ga_{0.24}In_{0.52}As$}}
\end{picture}
\begin{bf}
\newpage
\centerline{Fig.2}
\vspace{0.5cm}
\centerline{(a)}
\vspace{0.4cm}
\begin{picture}(100,220)(-170,20)
\put(-8,240){\makebox(0,0){$Al_{o.24}Ga_{0.24}In_{0.52}As$}}
\put(0,220){\makebox(0,0){1.13 eV}}
\put(40, 215){\line(-1,0){80}}
\put(0,30){\makebox(0,0){-0.16 eV}}
\put(40, 20){\line(0,1){210}}
\put(80,210){\makebox(0,0){0.9604 eV}}
\put(40, 205){\line(1,0){80}}
\put(80,190){\makebox(0,0){0.765 eV}}
\put(80, 185){\makebox(0,0){$-\cdot-\cdot-\cdot-\cdot-\cdot-$}}
\put(80, 100){\makebox(0,0){-0.042 eV}}
\put(40, 95){\line(1,0){80}}
\put(80,175){\makebox(0,0){0.685 eV}}
\put(80, 170){\makebox(0,0){$\bullet\bullet\bullet\bullet\bullet\bullet\bullet\bullet\bullet\bullet\bullet$}}
\put(80,88){\makebox(0,0){-0.0436eV}}
\put(80, 83){\makebox(0,0){$-\cdot-\cdot-\cdot-\cdot-\cdot-$}}
\put(80,146){\makebox(0,0){0.0122 eV}}
\put(80, 140){\makebox(0,0){$\bullet\bullet\bullet\bullet\bullet\bullet\bullet\bullet\bullet\bullet\bullet$}}
\put(80,130){\makebox(0,0){0.009 eV}}
\put(80, 125){\makebox(0,0){$\bullet\bullet\bullet\bullet\bullet\bullet\bullet\bullet\bullet\bullet\bullet$}}
\put(80,112){\makebox(0,0){-0.023 eV}}
\put(80, 107){\makebox(0,0){$-\cdot-\cdot-\cdot-\cdot-\cdot-$}}
\put(40, 25){\line(-1,0){80}}
\put(80,240){\makebox(0,0){$Ga_{0.6}In_{0.4}As$}}
\put(120, 20){\line(0,1){210}}
\put(160,215){\makebox(0,0){1.0177 eV}}
\put(160, 210){\makebox(0,0){$-\cdot-\cdot-\cdot-\cdot-\cdot-$}}
\put(160,205){\makebox(0,0){0.948 eV}}
\put(160, 198){\makebox(0,0){$\bullet\bullet\bullet\bullet\bullet\bullet\bullet\bullet\bullet\bullet\bullet$}}
\put(160,81){\makebox(0,0){-0.0635 eV}}
\put(160, 75){\makebox(0,0){$\bullet\bullet\bullet\bullet\bullet\bullet\bullet\bullet\bullet\bullet\bullet$}}
\put(160,69){\makebox(0,0){-0.0766 eV}}
\put(160, 62){\makebox(0,0){$\bullet\bullet\bullet\bullet\bullet\bullet\bullet\bullet\bullet\bullet\bullet$}}
\put(160,54){\makebox(0,0){-0.0954 eV}}
\put(160, 47){\makebox(0,0){$-\cdot-\cdot-\cdot-\cdot-\cdot-$}}
\put(160,43){\makebox(0,0){-0.09942 eV}}
\put(160, 36){\makebox(0,0){$-\cdot-\cdot-\cdot-\cdot-\cdot-$}}
\put(160,185){\makebox(0,0){0.713 eV}}
\put(120, 180){\line(1,0){80}}
\put(160,105){\makebox(0,0){-0.028 eV}}
\put(120, 100){\line(1,0){80}}
\put(180,240){\makebox(0,0){$Ga_{0.4}In_{0.6}As$}}
\end{picture}

\vspace{1cm}
\centerline{(b) }
\vspace{0.4cm}
\begin{picture}(100,220)(-175,20)
\put(-8,240){\makebox(0,0){$Al_{o.24}Ga_{0.24}In_{0.52}As$}}
\put(0,220){\makebox(0,0){1.13 eV}}
\put(40, 215){\line(-1,0){80}}
\put(0,30){\makebox(0,0){-0.16 eV}}
\put(40, 20){\line(0,1){210}}
\put(80,195){\makebox(0,0){1.0827 eV}}
\put(40, 190){\line(1,0){80}}
\put(80,185){\makebox(0,0){0.634 eV}}
\put(80, 180){\makebox(0,0){$-\cdot-\cdot-\cdot-\cdot-\cdot-$}}
\put(80, 100){\makebox(0,0){-0.049 eV}}
\put(40, 95){\line(1,0){80}}
\put(80,175){\makebox(0,0){0.541 eV}}
\put(80, 170){\makebox(0,0){$\bullet\bullet\bullet\bullet\bullet\bullet\bullet\bullet\bullet\bullet\bullet$}}
\put(80,88){\makebox(0,0){-0.014 eV}}
\put(80, 83){\makebox(0,0){$-\cdot-\cdot-\cdot-\cdot-\cdot-$}}
\put(80,146){\makebox(0,0){0.0523 eV}}
\put(80, 140){\makebox(0,0){$\bullet\bullet\bullet\bullet\bullet\bullet\bullet\bullet\bullet\bullet\bullet$}}
\put(80,130){\makebox(0,0){0.0339 eV}}
\put(80, 125){\makebox(0,0){$\bullet\bullet\bullet\bullet\bullet\bullet\bullet\bullet\bullet\bullet\bullet$}}
\put(80,112){\makebox(0,0){0.0112 eV}}
\put(80, 107){\makebox(0,0){$-\cdot-\cdot-\cdot-\cdot-\cdot-$}}
\put(40, 25){\line(-1,0){80}}
\put(80,240){\makebox(0,0){$Ga_{0.7}In_{0.3}As$}}
\put(120, 20){\line(0,1){210}}
\put(160,215){\makebox(0,0){1.116 eV}}
\put(160, 210){\makebox(0,0){$-\cdot-\cdot-\cdot-\cdot-\cdot-$}}
\put(160,205){\makebox(0,0){1.115 eV}}
\put(160, 200){\makebox(0,0){$\bullet\bullet\bullet\bullet\bullet\bullet\bullet\bullet\bullet\bullet\bullet$}}
\put(160,81){\makebox(0,0){-0.0986 eV}}
\put(160, 75){\makebox(0,0){$\bullet\bullet\bullet\bullet\bullet\bullet\bullet\bullet\bullet\bullet\bullet$}}
\put(160,69){\makebox(0,0){-0.117 eV}}
\put(160, 62){\makebox(0,0){$\bullet\bullet\bullet\bullet\bullet\bullet\bullet\bullet\bullet\bullet\bullet$}}
\put(160,54){\makebox(0,0){-0.123 eV}}
\put(160, 47){\makebox(0,0){$-\cdot-\cdot-\cdot-\cdot-\cdot-$}}
\put(160,43){\makebox(0,0){-0.131 eV}}
\put(160, 36){\makebox(0,0){$-\cdot-\cdot-\cdot-\cdot-\cdot-$}}
\put(160,190){\makebox(0,0){0.6427 eV}}
\put(120, 185){\line(1,0){80}}
\put(160,105){\makebox(0,0){-0.021 eV}}
\put(120, 100){\line(1,0){80}}
\put(180,240){\makebox(0,0){$Ga_{0.4}In_{0.6}As$}}
\end{picture}
\end{bf}
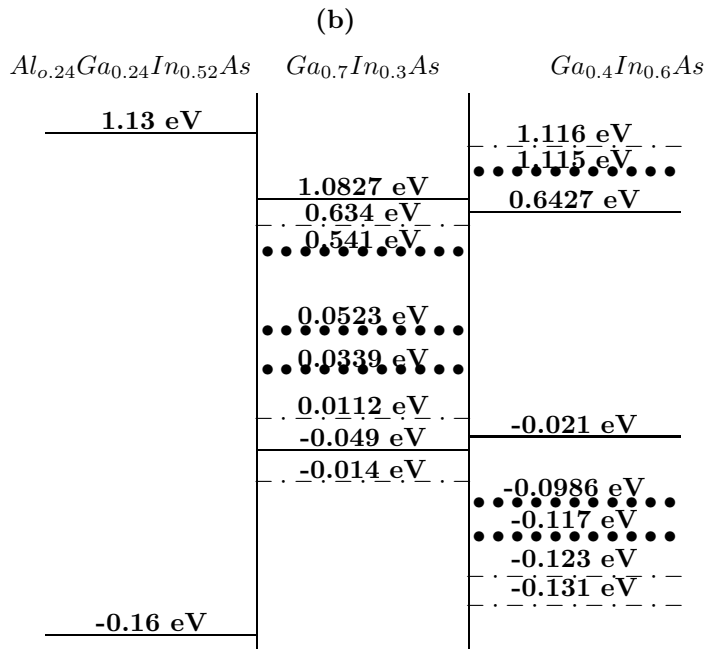
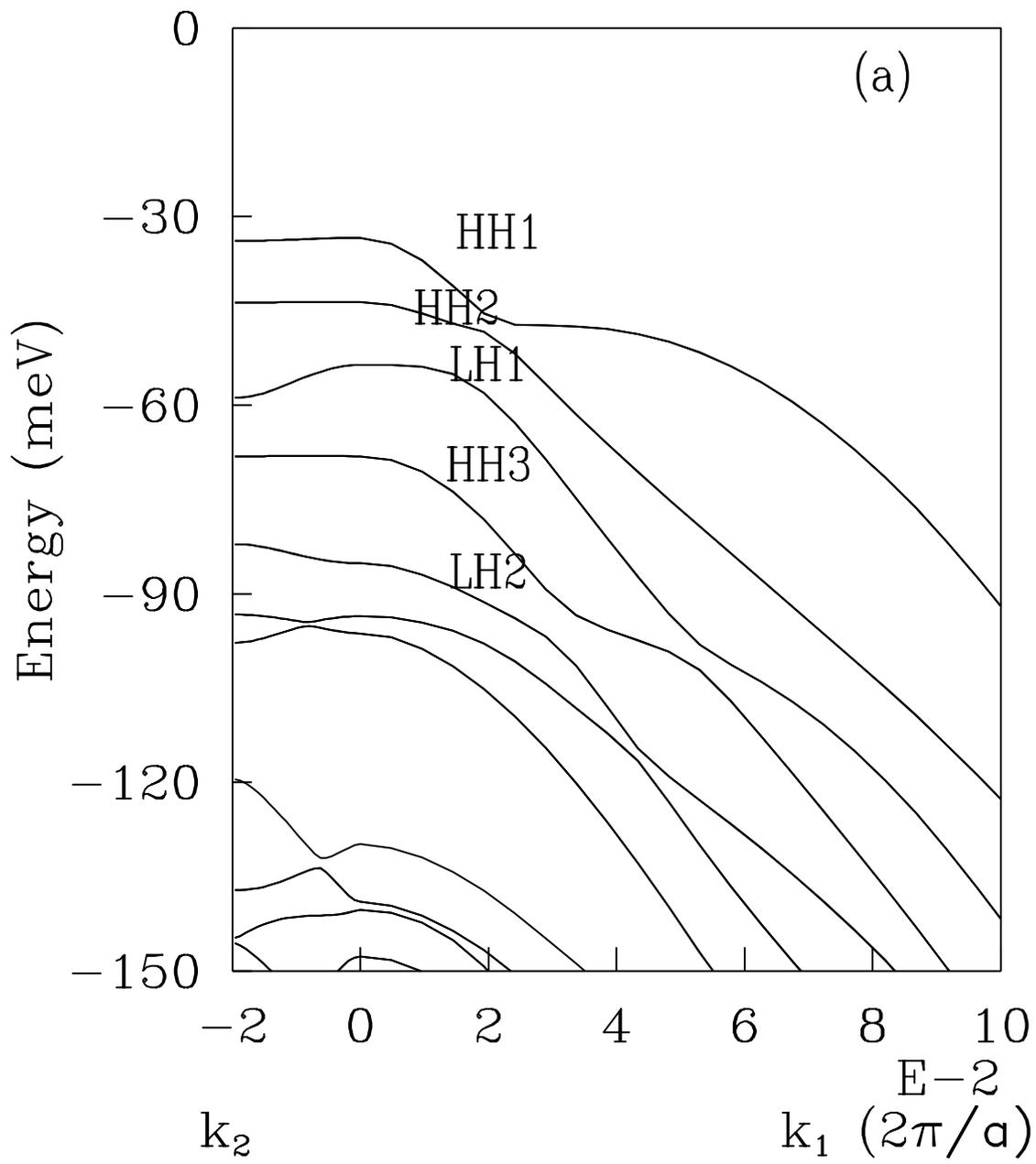
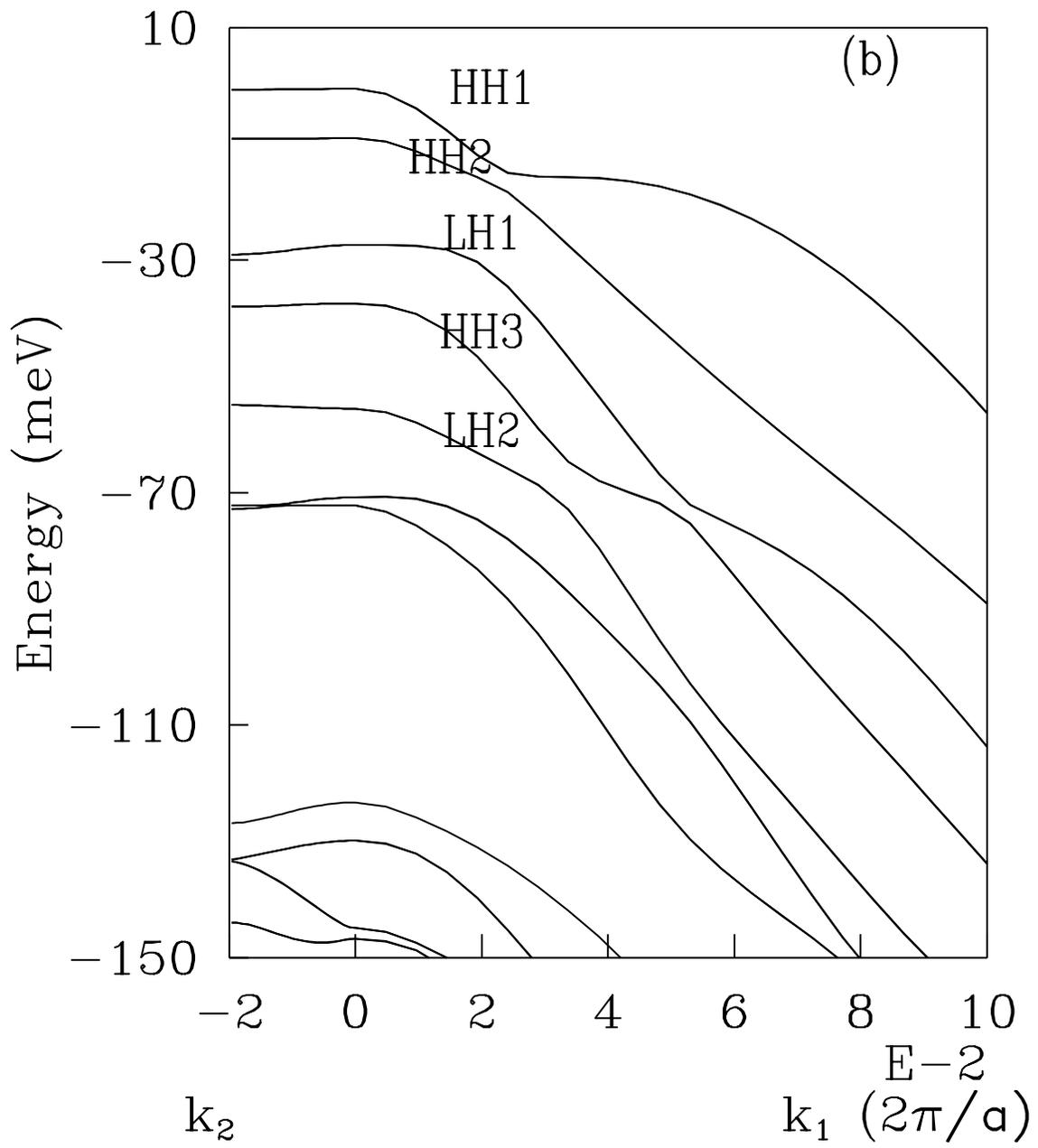
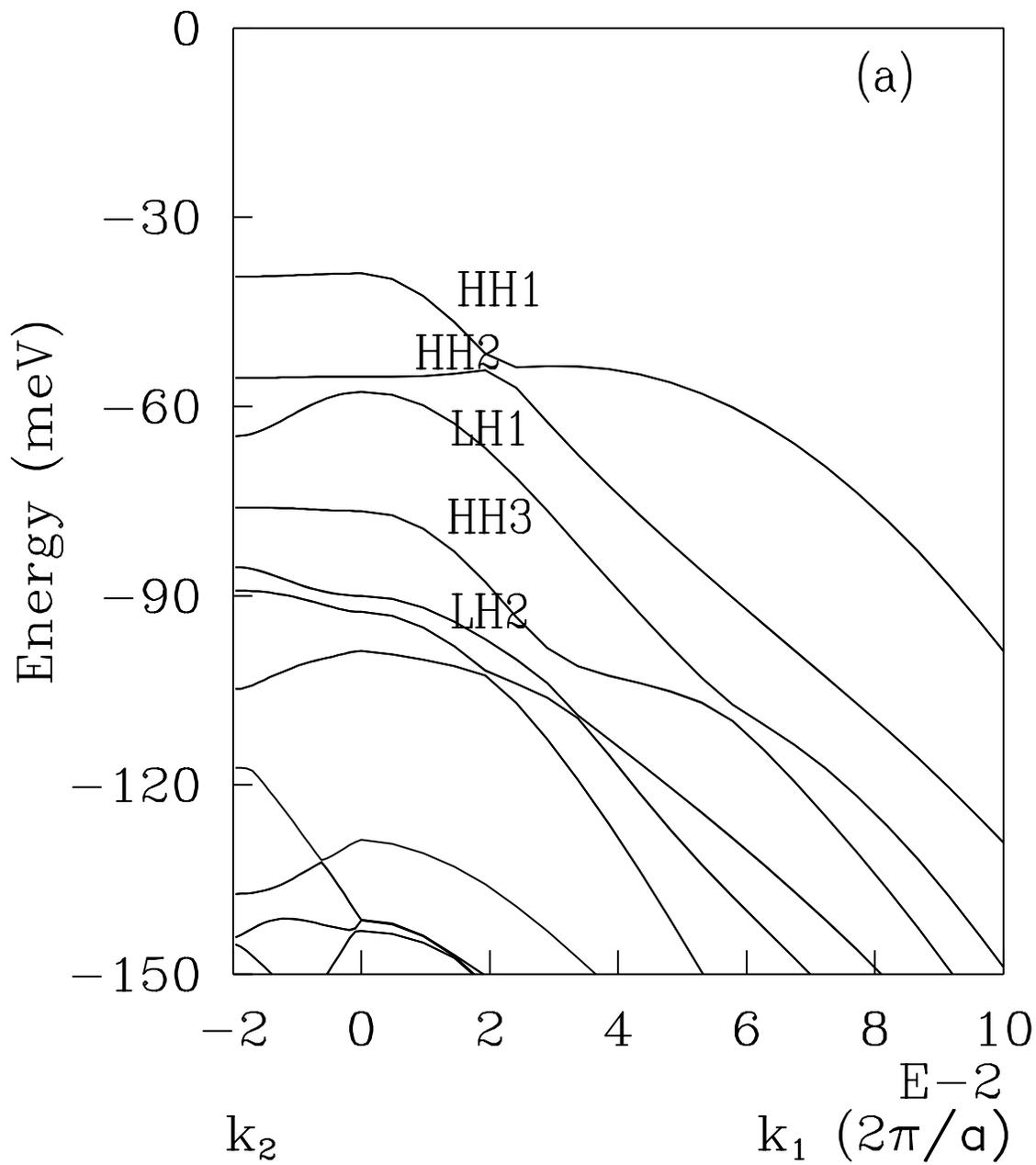
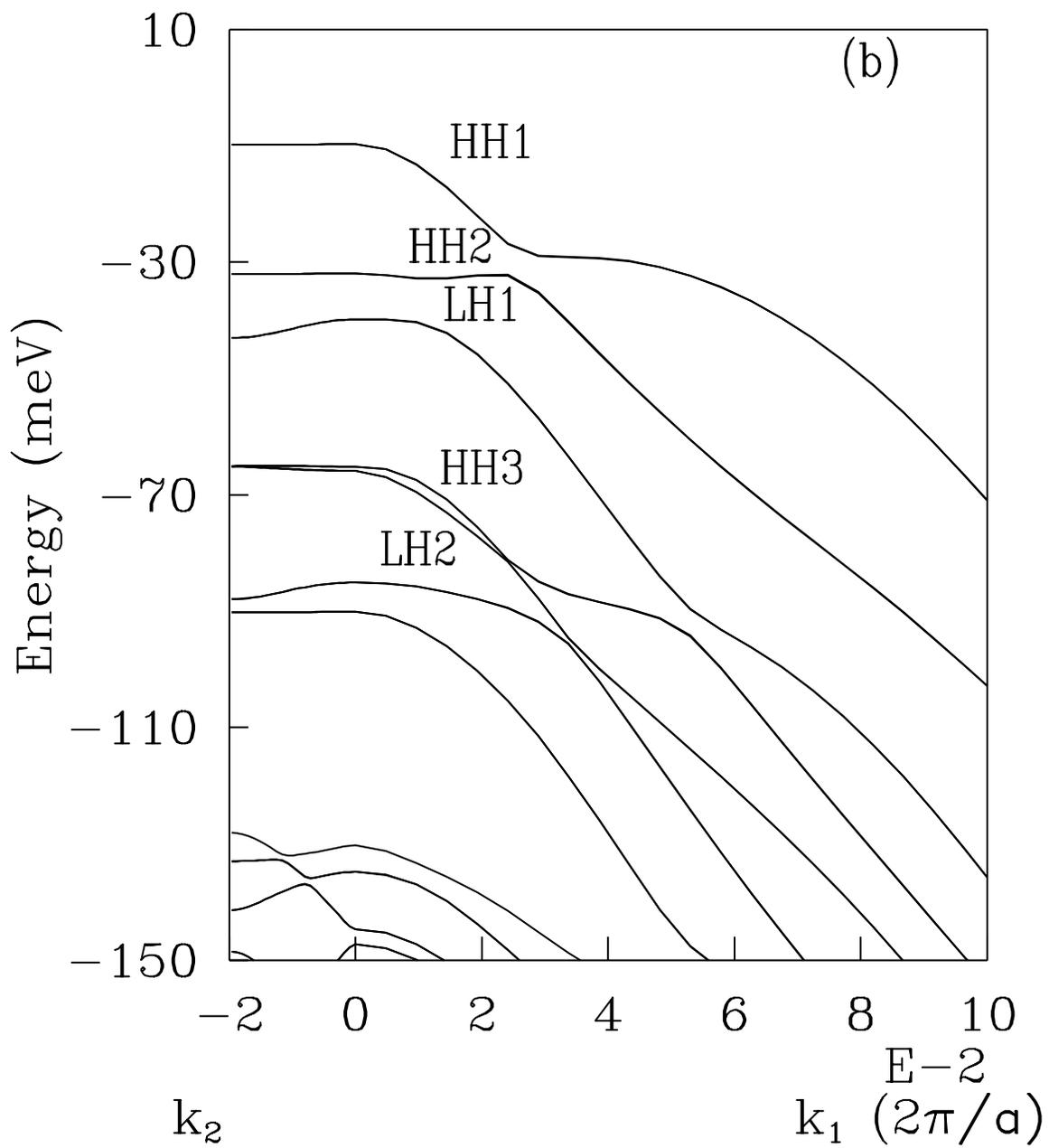
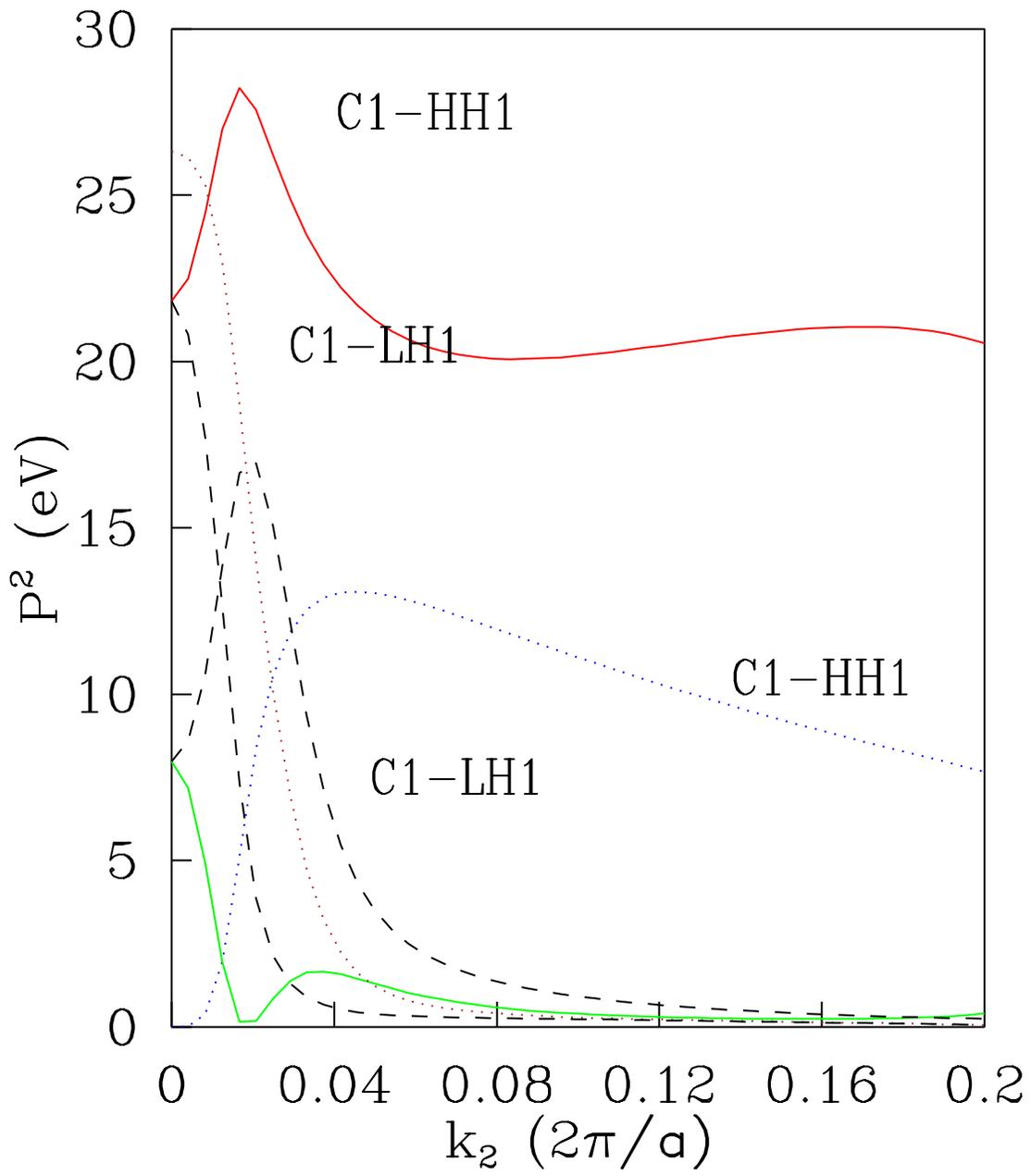
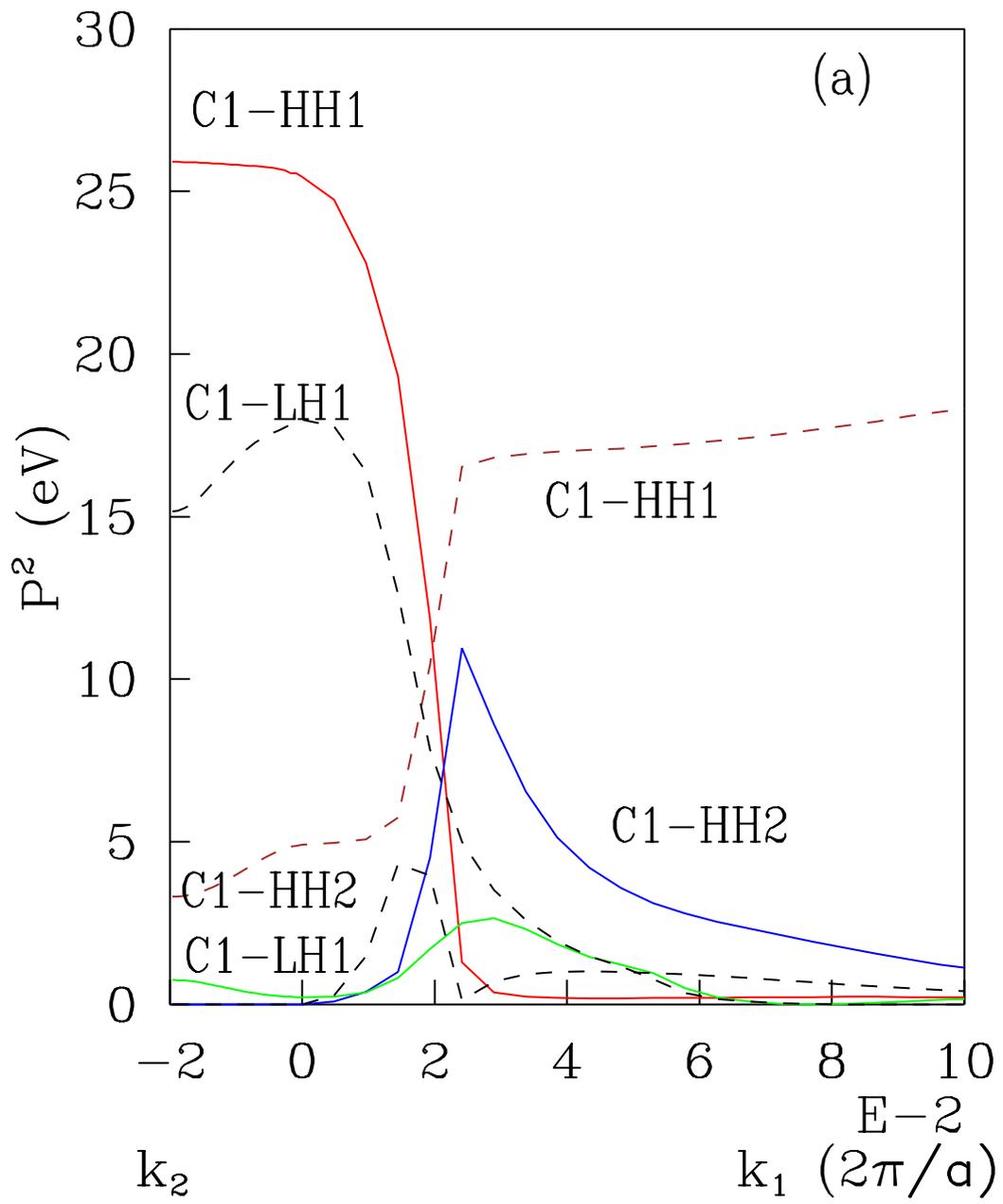
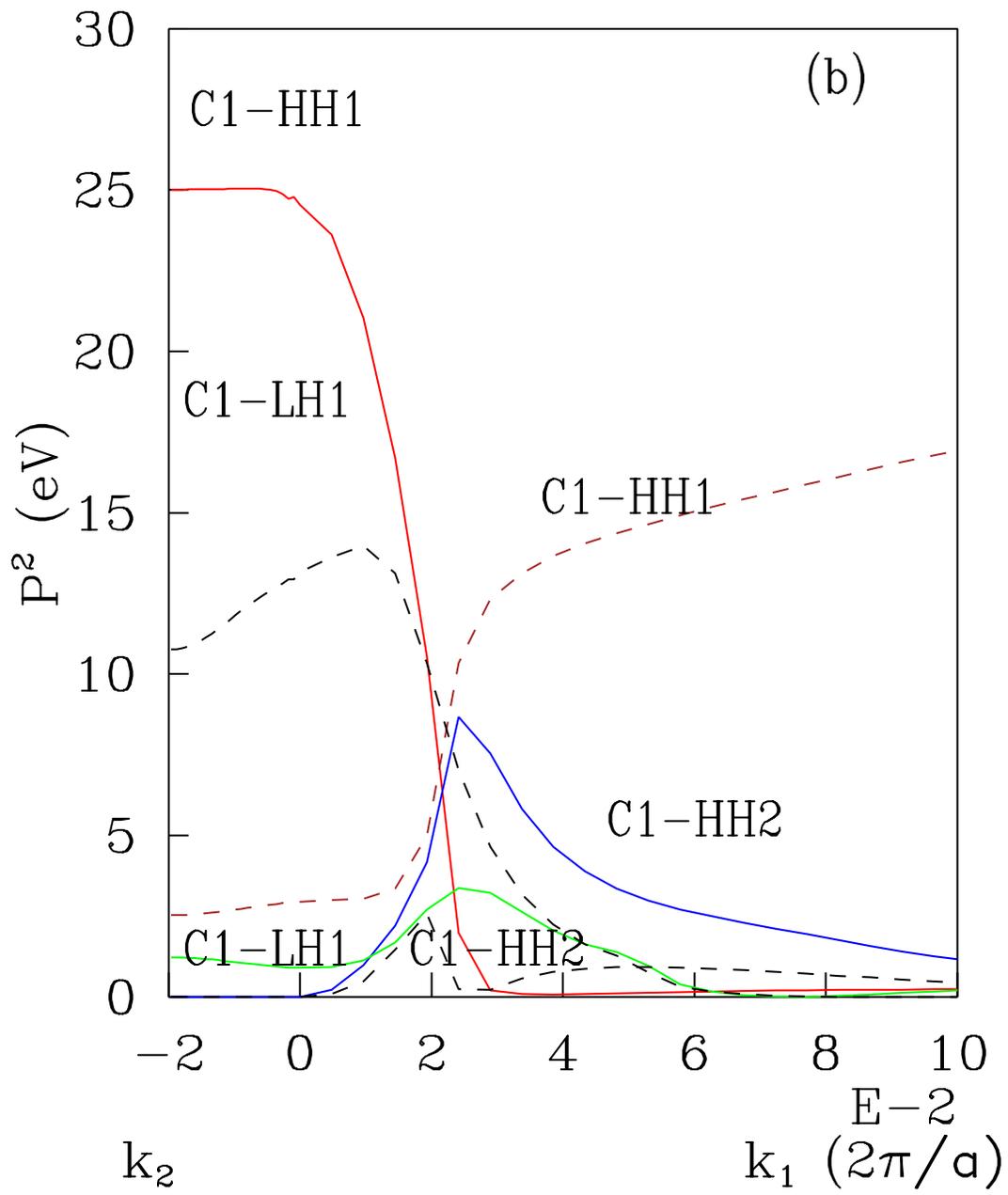
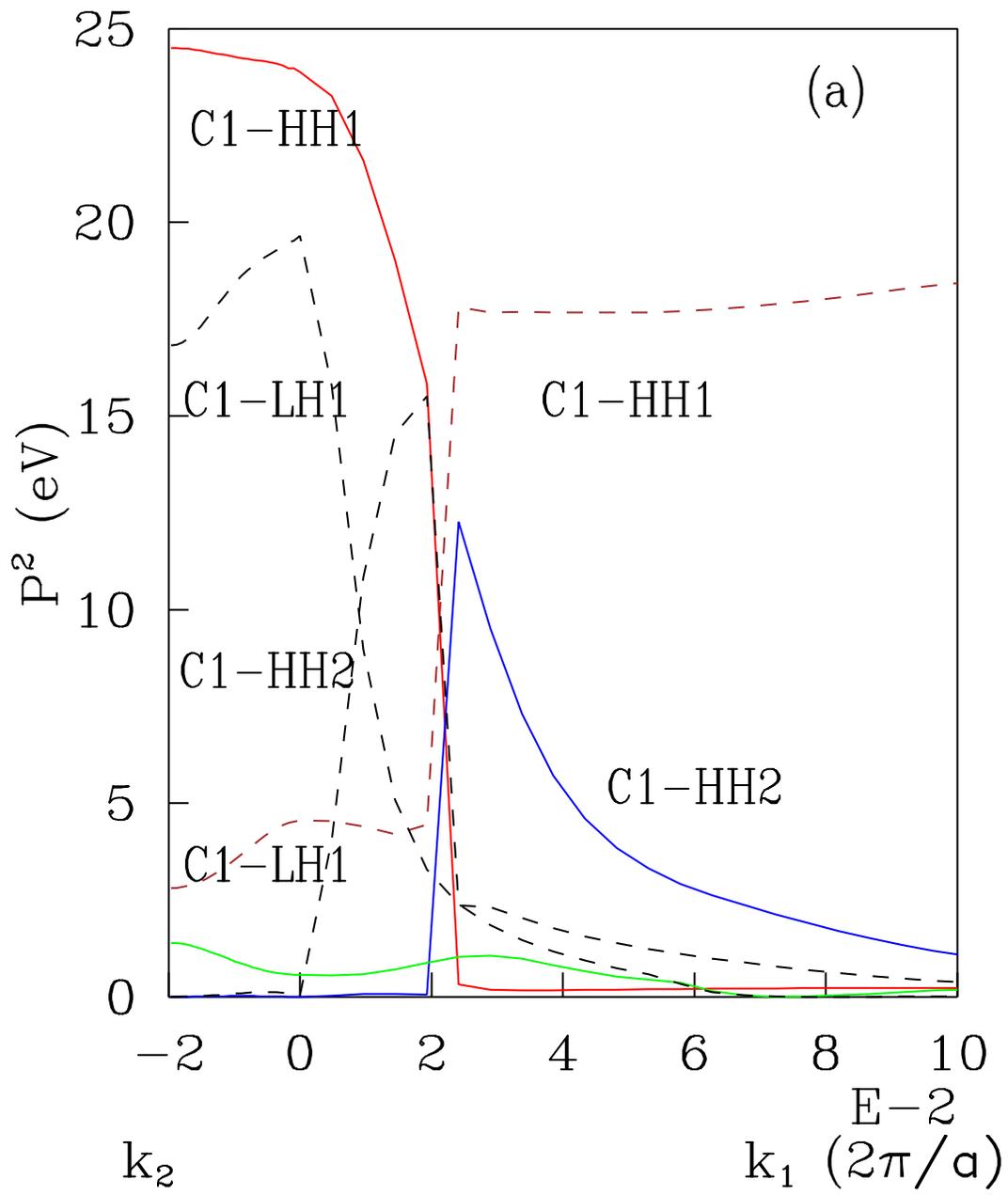
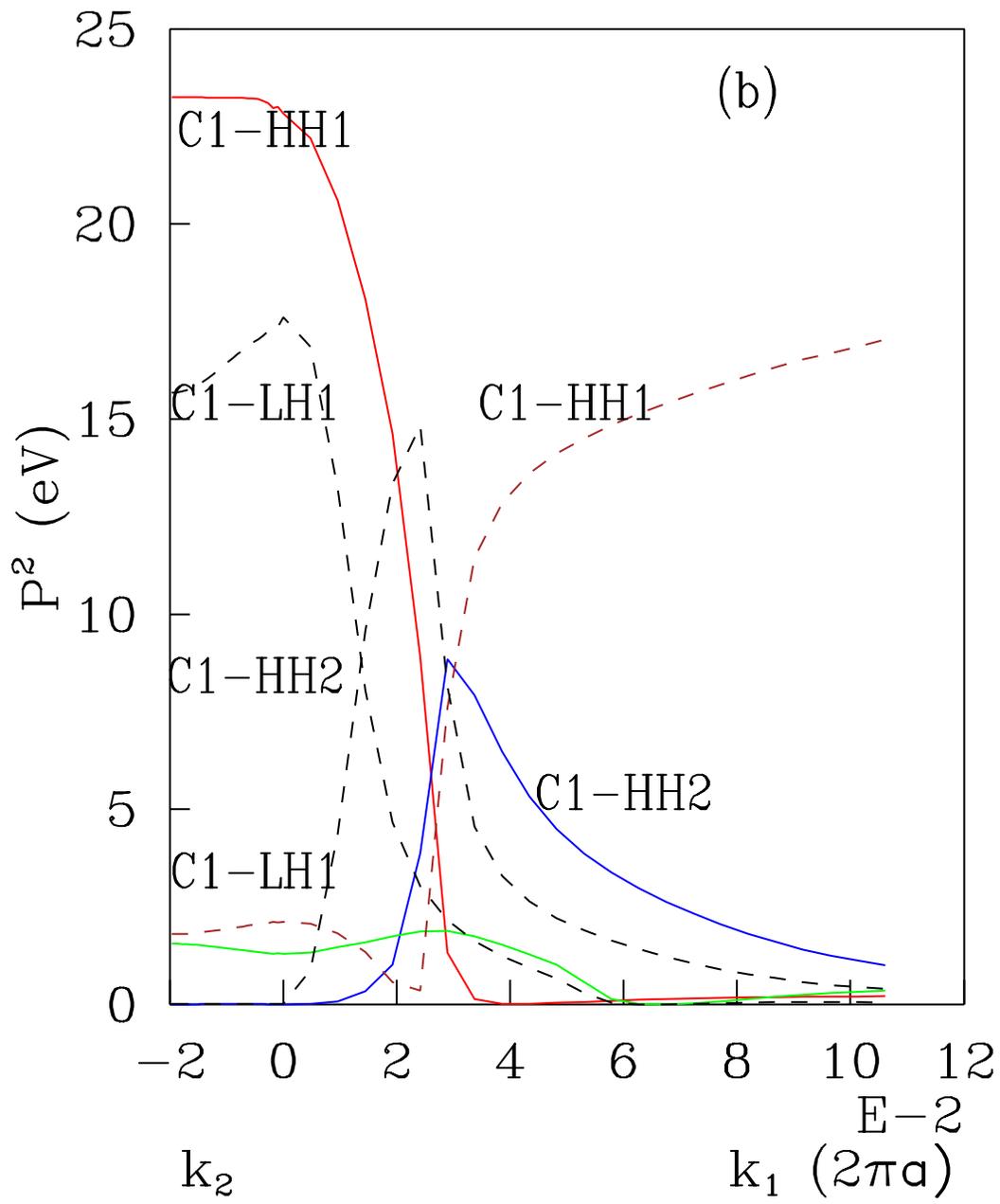
\end{document}